\documentclass[runningheads]{llncs}
\usepackage{graphicx}
\newcommand{\methodName}{Find It}
\begin{document}
\title{Find It: A Novel  Way to Learn Through Play}
%
%
\author{Md. Tashfiqul Bari, Tanvir Hassan, Raisa Tabassum, Zubaida Ahmed and Swakkhar Shatabda}
\authorrunning{Bari et al.}
%
\institute{Department of Computer Science and Engineering, United International University, Plot 2, United City, Madani Avenue, Badda, Dhaka-1212, Bangladesh.
\email{tashfiqulbari@gmail.com,tanvirhassan47@gmail.com,raisa08bd@gmail.com, zubaidazubu4@gmail.com,swakkhar@cse.uiu.ac.bd}}
\maketitle              
\begin{abstract}
Autism Spectrum Disorder (ASD) is the area where many researches enduring like Magnetic Resonance Imaging (MRI), called diffusion tensor imaging, Early Start Denver Model (ESDM) to provide an easier life for the people diagnosed. After years and years of combined funding sources from public and private funding, these researches show great promises in recent years. In this paper, we have tried to show a way how children with Down Syndrome Autism can learn through game therapy. These game therapies have shown an immense number of improvements among those children to learn alphabets along with developing their motor skills and memory challenges.

\keywords{Down Syndrome \and Serious Games \and Learning Analytics.}
\end{abstract}

\section{Introduction}
Autism is a diffusive developmental disorder in early childhood. As parents none wants their children to have any problem but when it comes to an autism, people should give serious attention. Children with autism face immense struggles when it comes interacting with their typically developing peers and also in their learning process. It should be carefully handled both their mental and physical health in every situation. Many individuals figure children with a mental imbalance ought to invest less energy playing contrasted with non-impaired children. In reality, we should pay more attention to autistic children \cite{item1}.
	
Without a doubt ASD isn't something a child just `grows out of', there are numerous medications that can enable to secure new aptitudes and overcome a wide variety of formative difficulties. From free government organizations to in-home behavioral treatment and school-based projects, help is accessible to meet child’s unique needs. With the correct treatment design, and a considerable measure of affection and support, child can learn, develop, and thrive. The most punctual indications of extreme introvert are the nonappearance of typical practices and the nearness of strange ones so they can be difficult to perceive. The earliest symptoms of autism children are as signs calm, free, and undemanding. Despite the fact that a mental imbalance is difficult to analyze before 24 months, symptoms regularly surface in the surrounding area of 12 and year and a half. If signs are distinguished by year and a half of age, concentrated treatment may rewire the mind and turn around the symptoms. Children with ASD hang loose applying what they have taken in one session from a specialist, home or others. Having a consistent way of interaction with a special child and it should stick to a schedule which is best for them. Positive behavior can go long way with children. We should praise them what they act appropriately or learn new skill and give some reward for their performances. Creating a comfort zone at home helps them a lot. In research of ten years, it has been seen that computer-based interventions can provide extra ordinary strategies to help the children with special needs in many ways.

Autism is the most commonly found neuro-development disorder and its core deficits in three domains: social interaction, communication, and repetitive or stereotypic behavior. It’s been calculated that 1$\%$  of the world’s population, suffer from an autism spectrum disorder. In many developing countries like Bangladesh also have no data to measure the summation that how many children or adults are suffering from this lifelong neurological condition. In  recent years, Bangladesh has been developing facilities for special children with autism \cite{item3,item4}.

In this paper, we address the issue of improving learning capacity through our designed applications \cite{item2}. Our examination has focused on smart phone-based applications to improve the learning skill of children with Down syndrome through playing some simple mini games in several levels.

\section{Understanding Down Syndrome}

Human body is nothing less than a miracle. How human body will develop, look and work depends on the fact known as “genes”.  Genes are the reason behind every characteristic of a human body. They are also responsible for any abnormalities in a human body. People by birth have 22 chromosomes but people with Down Syndrome are born with 23 chromosomes in their bodies. Chromosomes are the set of genes, with Down syndrome, this extra chromosome causes issues that affects them for life time.
	
Although Down Syndrome is a lifelong problem but with the help of modern science and treatments, at present doctors are helping these patients. With proper care and education, Down syndrome issue can get a better solution. 

\subsection{Effects of Down Syndrome}

Down syndrome occurs in about one per 1,000 babies born each year all over the world. It varies in characteristic and occurs differently in people. Some may suffer from understanding where some may suffer from interaction with surroundings. With proper care, these issues can be handled. People with Down syndrome have some physical features in common. For example, flat noses, small ears, straight hair etc. They’ll learn skills gradually but will face problems in daily activities like walking, talking, and developing social skills.

\subsection{Causes of Down Syndrome}

The worldwide accepted main reason behind down syndrome is to have one extra chromosome in human body. There is a higher chance that women aged 35 and older and already having a child with Down syndrome, are more likely to have another one who has it as well. 

It is possible to pass Down syndrome from parent to child. Again, parents having no Down syndrome can have down syndrome child because they have correct number of genes, but their child doesn’t.

\subsection{Life of a Down Syndrome child}

Individuals with Down syndrome usually have cognitive development profiles that suggest mild to moderate intellectual disability. However, cognitive development and intellectual ability are highly variable. Children with Down syndrome often reach developmental milestones later than their peers. There may be a delay in acquiring speech. A child may need speech therapy to help them gain expressive language. Fine motor skills may also be delayed. They can take time to develop after gross motor skills have been acquired.\\
	
	On average, a child with Down syndrome will sit at 11 months, crawl at 17 months and walk at 26 months. There may also be problems with attention, a tendency to make poor judgments, and impulsive behavior. However, people with Down syndrome can attend school and become active, working members of the community.\\
	
	Sometimes, there are general health problems that can affect any organ system or bodily function. Around half of all people with Down syndrome have a congenital heart defect.

	There may also be a higher risk of: respiratory problems, Hearing difficulties, Alzheimer's disease, childhood leukemia, epilepsy and thyroid conditions. However, there also appears to be a lower risk of hardening of the arteries, diabetic retinopathy and most kinds of cancer.

\subsection{Sign and Syndromes}

There are various types of a person with Down Syndrome autism and there are many forms of disorder which cannot be described easily \cite{item5}.
	
	It’s easy to fall into thinking that everyone with Down Syndrome looks similar but in reality, Down syndrome affects people both physically and mentally very differently in each of them. And there’s no prediction how it will affect anybody in the long run. For some of them, the effects are normal, they can even have an easier life style. But for others it is impossible to do daily activities without the assist of anyone else.

\subsubsection{Physical and Mental Symptoms}

Some common physical features that are seen in down syndrome children are flatter faces, almonds shaped eyes, small ears, small hands and feet, short neck and small head.

Some common mental symptoms that are seen in down syndrome children are: they may suffer from hearing loss, they may have heart problems, suffer from obstructive sleep apnea, most common eye sight problems and develop several blood conditions and infections.	

\subsection{Treatment Strategies}
\subsubsection{Applied Behavior Analysis (ABA)}

Applied Behavior Analysis (ABA) is the direction of learning and encouragement from Behavior Analysis, and the strategies and innovation got from those standards to the arrangement of issues of social centrality. The thought process of this technique is advance. This technique enhances such huge numbers of issues, for example, correspondence, creative energy, discretion, self-observing. It is organized and regular practice process which helps to enhance over the long haul. A child can response in three ways of any method. Behind every child, they need individual therapy which is given by the instructor. So many dynamic benefit of a package of ABA methods in Comprehensive, Individualized and Intensive Early Intervention projects for special children. Comprehensive mention the skill of day to day life for children and self-control and motivation. Early intervention designed the entire beginning program before age 4. Intensive refers to the total program per week 30-40 hours for 1-3 years children \cite{item6}.

\subsubsection{Discrete Trial Training (DTT)}

Discrete trial preparing (DTT) is a strategy for instructing in which the grown-up utilizes grown-up coordinated, fortifies decided for their quality, and clear possibilities and duplication to educate new abilities. DTT is a particularly solid technique for inspiring for another reaction to an incitement. Its restrictions are absence of fortification of student immediacy. Using DTT this learner follows some steps. In 1st step teacher or instructor decide for learners what are the objectives can be taught using DTT and summarize the results. 2ND step teacher completes the student analysis task and lists what to do and how a student can do that task. 3rd step is about Setting up the Data Collection by teachers. In 4th mentors designed the location which can take place. 5th is teacher gather the materials which will be used in practice. In 6th teacher assists the learner and gather all the attention. In 7th massed the trial of the learner. 8th is what teacher conducting with the learner and lastly means in 9th mentors give a review and modify the results \cite{item7}.

\subsubsection{Functional Communication Training (FCT)}

Functional communication training (FCT) is an effective finding of behavior problems. This method described by Carr and Durand (1985), differential reinforcement is used for this method. People who have lots of aggression, people who hurt themselves, vocal problem, stereotypy participants are applicable for this therapy. A variety of response targeted in FCT, including vocal responses, picture exchanges, sign language, gestures, and activation of voice or text output devices \cite{item8}.

\subsubsection{Incidental Teaching}

Incidental teaching creates an environment for special children which introduce children interest. This method used to add fun to children life. Six principle serves this method. First is early intervention is essential and proper time to develop this method between the ages 15 to 30 months. Second is the improvement is not going to show you in one day it’s just they should engage with minimum 30 hours per week. The third is home and company with parents is most important of upbringing them. 4th is they should interact with socially and for that need to plan how can they interact. 5th is children need to learn to speak in a discrete trial. And final is depending upon incidental teaching procedures within the environment \cite{item9}.

\section{Related Work}

In past decades, many researches are conducted on the issue of autism and how to solve this issue with the help of technology. On the internet we can find numerous serious games, casual games and program are made to help the autistic children. Hanan M. Zakari analyzed a huge amount of these games and tried to categorize them based on their primary goal \cite{item10}. This category helped us to define that specific sector, Down Syndrome we want to improve. From our research, not all of these games are highly effective. In fact, a little of them are effective in real life scenario.
	
One of the most common problem with Down Syndrome children are delayed language and speech development. Some children cannot pronounce the word or sentence correctly and find trouble to express their thoughts. Rahman and his team tried to solve this problem by playing an e-learning iterative game to the children. The teacher will choose some picture of different things and pass those pictures through LAN. The autistic children will receive them in their computers. These pictures will show at the left of the screen and begin to move to right of the screen. In this time, the children will pronounce the meaning of those pictures loudly and clearly so that the system can record that. Now the system will analyze the speech and if it is correct, that child will get the score \cite{item11}.
	
Related to our previous reference, another interesting game was made to help children speak properly. Many autistic children speak so fast that other person can not understand them. To control their speaking, M. Goodwin and his team created a turtle race serious game. Children have to speak in proper speed, loudness and clearness to win the race. They can not speak very fast nor very slow. To win the race, they have to maintain the moderate speed of speaking \cite{item12}
	
Another major difficulty autistic children face is the poor emotion recognition of other people. LIFEisGAME is a serious game that help children to understand facial expression of a person in different levels. Tech giant Microsoft formed a group with specialist from Portugal and University of Texas to develop this game. This game have multiple stage. The very first one is the $``$Recognize the expression$"$ . A 3D cartoon model will be shown in the screen and children have to recognize the emotion of that model by reading its facial expression. In the next stage $``$Build a Face$"$ , children have to rebuild a face on the basis of emotions on cards given. The final stage of this game is $``$Live the Story$"$ , where the children will play a sort story of a cartoon character. The main goal of this story is to express emotions correctly in certain scenarios \cite{item13,item14}.
	
Modern technology opened a door to solve the problems of autism more efficiently. Virtual Reality (VR) is one of the most promising among of these technologies that can help us greatly to solve many of these problems. Marco\textit{ }Simões and his team develop an interesting game to help not only the children but also the young adults. With the help of VR, they created a game that will help ASD patient to give the experience of bus journey. They staged a small scenario where the player have to complete a bus journey. Player have to perform some small tasks like buying the ticket, waiting for the bus, choosing a sit in the bus, get down on the stoppage etc. With the VR headset, this game can give them almost the real feel of a true bus journey. This will help them to deal with the real life scenarios \cite{item15}.

\section{Our Method}
{\methodName} is a cross platform serious game that has been developed with an intension of teaching its user in a fun way to develop their certain skill set. The game has been developed for both Android and iOS platforms to reach the maximum user in the industry.

\subsection{Game Architecture}
{\methodName}  is a mobile application that has been specially designed and developed for Autistic children. The game consists of two mini game. The game `Balloon Pop'  will help to develop your skill for fast letter recognition while `Match Making'  will help you by challenging your memory to solve the different board puzzle. Our adaptive learning AI, will challenge the user based on their performance. While their performance gets stored to our Firebase cloud database for our further research and performance analysis. Involving user to an environment that is comfortable to explore with the learning material ensuring that player is not always engaged with the less performance game, but also engaging with the most performing game in order to keep the $``$fun vs. challenge$"$  curve consistent.

\begin{figure}
	\begin{center}
	\includegraphics[width=0.5\textwidth]{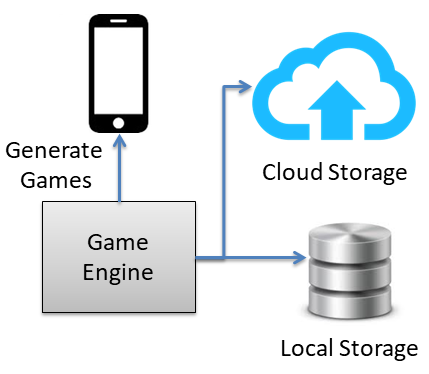}
\end{center}
\caption{Basic Game Architecture. \label{figArch}}
\end{figure}

\begin{figure}
	\centering
	\includegraphics[width=\textwidth]{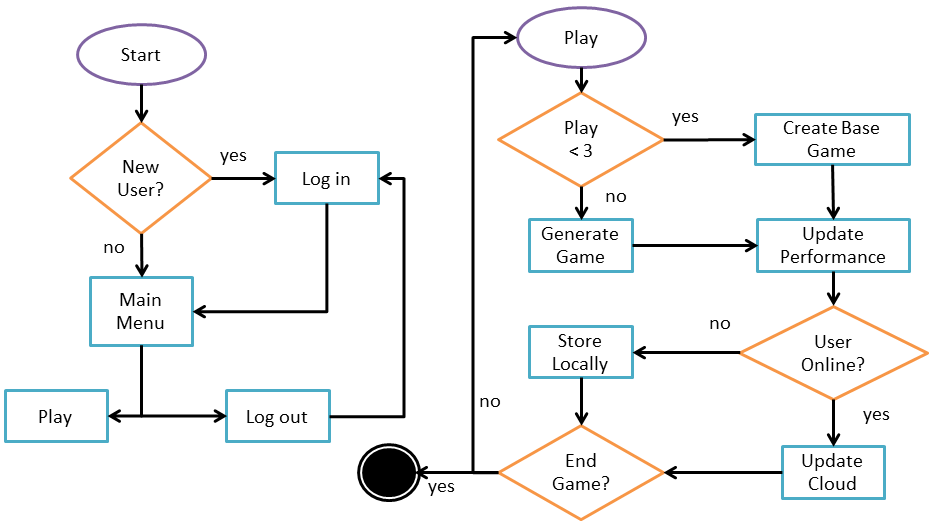}
\caption{A simplified flowchart of the game. \label{figFlow}}
\end{figure}

At the center of the game architecture is a game engine that is in the control of three processes: i) game generation, ii) storage of information and iii) user management. The basic architecture of {\methodName} is shown in Figure~\ref{figArch}. The basic game generation is based on learning analytics and it keeps user records via online and local storage. A new user willing to play the game has to log in to the system. A user who have not played the game that much frequently, his or her games are generated based on a basic game play module. However, playing each game improves the performance and the analytics is applied after at least three game plays. After each play, the performance is stored in cloud if the user is online or the information is locally stored when offline with an option to be synced with the online storage when the user is online again. The analytics generates game each time after that minimum number of plays based on their previous performances. A flow chart of the game engine is shown in Figre~\ref{figFlow}. 
\subsection{The Gameplay}

\begin{figure}
	\centering
	\begin{tabular}{ccc}
		\includegraphics[width=0.3\textwidth]{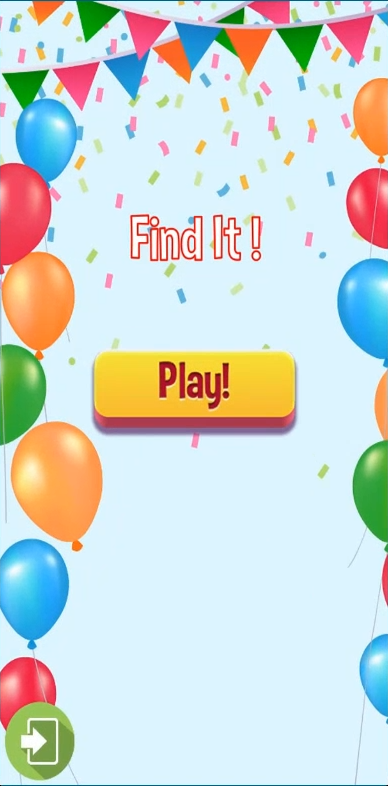}&\includegraphics[width=0.3\textwidth]{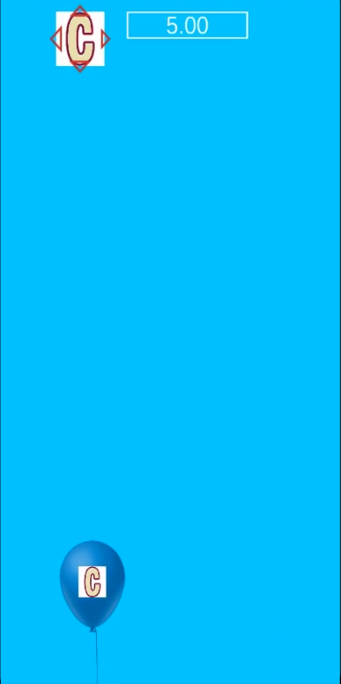}&\includegraphics[width=0.3\textwidth]{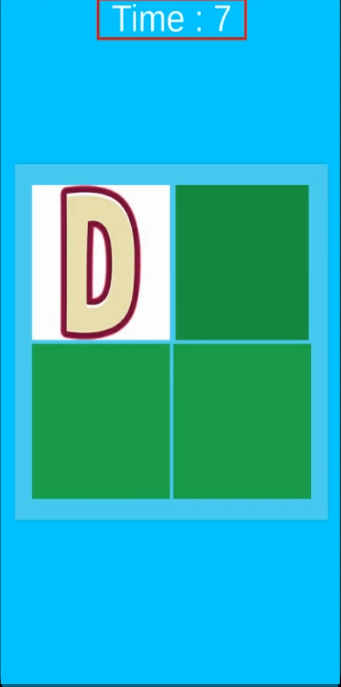}\\
		(a)&(b)&(c)\\
	\end{tabular}
\caption{Screenshots of the game play: (a) main menu, (b) balloon pop and (c) match making. \label{figPlay}}
\end{figure}

After Installing and opening the application, user have to log in by the email ID in order to get track on their performance report. The user can logout or switch between accounts with the different email. After the login process complete, user simply have to tap on the $``$Play Button$"$. The game will start with one of the mini game that has been available to play. If the game is $``$Balloon Pop$"$ , the user has to tap on the balloon with the targeted $``$Letter$"$  that has been displayed as the target. If the targeted letter is $``$A$"$ , the user has to tap on the balloon consist of letter $``$A$"$  and avoid tapping on other balloon. At the end of $``$Balloon Pop$"$  game, the performance will be stored in firebase cloud database based which is based on accuracy. If the game is $``$Matchmaking$"$ , the user has to match the two same card with the same letter in order to complete the board. So if the targeted letter is $``$A$"$ , user has to match all pair of $``$A$"$  at the board in order to complete the level. At the end of $``$Matchmaking$"$  game. The performance will be stored in firebase cloud database which is based on time. Screenshots of the gameplays are shown in Figure~\ref{figPlay}.

\subsection{Implementation}
We used $``$Unity$"$  as cross platform game engine to develop our game to reach out the maximum user on different platform. We used $``$Firebase$"$  to store our user data which handle the both offline $\&$  online situation of our application. We also have deployed $``$Vuforia$"$ , an $``$AR$"$  sdk for both $``$Android$"$  $\&$  $``$iOS$"$  platform

\section{Results and Discussion}
We have experimented with our game with special permission from the authorities of Smiling Special Children School for Special children at Badda, Dhaka. We gathered game play experience with the autistic children and their mentors. We made sure to maintain the ethical and privacy related issues while experimenting with our application and the survey. Ten mentors or teachers from the special school were involved in the survey part. Seven of them were female and three of them were male teachers. Their age was between 21-40 and 8 of them actually under thirty years of age. The goal of our session with the school was twofold. One is to get the children play the game using the app so that we can gather real time feedback. Also we wanted the mentors to rate the application and their experiences with the children.

The survey contained several questions for the teachers and mentors. There were three groups of questions: i) experience related, ii) responsibility related and iii) procedure related. In the experience related section there were four questions enlisted as below:
\begin{enumerate}
	\item How many years you have been taught autism children?
	\item What strategies/materials do you use when teaching students with autism?
	\item In what environmental setting you have most often teach?
	\item What subject area(s) you are currently teaching?
\end{enumerate}
In response to these  questions, we found that 60$\%$  among the participants are teaching around 4 to 6 years and 30$\%$  participants are teaching around 1 to 3 years and also 10$\%$  participants are teaching around 7 to 9 years. It is shown that most of the teachers are teaching around 1-6 years. In conclusion, they have significant experience on teaching autistic children. All of them agreed to provide practical treatment so that the children can develop them easily. The third question was multiple choice with options like traditional classroom, virtual classroom, playground and others. Among all participants 100\% of them prefer virtual classroom, general educational
classroom and playground which is most required as their preferences. In other section 30\% of them prefer outing program which is included shopping, picnic, tour etc. Most of the teachers were engaged in teaching mathematics and science, while a few in dancing, pre-writing, nonverbal act and dotted training to write alphabet.

The second section was on responsibility and following questions were asked:
\begin{enumerate}
	\item With	whom the children share good relationship?
	\item Number of
	students you are responsible for in a typical day?
	\item Is
	there any specific talent in children?
\end{enumerate}

Among all the participants 80\% participants answered that
children relationship is very close with parents and teachers , 10\% participants said
that children relationship with other autism children is also good. 10\% of children
is nearly happy with everyone and 10\% participants said that children do not share
good relationship with none of them and they also attention seeker. Among all of the participants 90\% participants said that
they are responsible for around 1-25 autistic children and 10\% of participants said
that 26-50 children. In most of the cases, they are engaged in one-to-one teaching. Among the participants of teachers answered that 70\% of
the children have some specific talent and 30\%of the children have no specific talent.

In the last section on teaching procedure, everyone were asked if they have any formal training on autism and on the scale of the effect of autism changing a students life. Among all the participants 50\% of teachers answered that
autism has huge impact on student life because some children can’t even spent
day to day life without any help. 40\% of participants answered that there are
some impact because those autism children can lead day to day life but they have
problem to learn new things. 10\% of participants answered no impact because
those children doing well and they are improving. Among all participants, 100\% of participants read book
and article about autism. 90\% of participants have completed autism-specific undergraduate
class. About 50\% of participants surfed internet to gather knowledge, 20\%
of participants have taken organizational training under that school where taught.
And 10\% of participants taken sports training from another institution.

We were also engaged with the mentors after the game playing session of the children. The game was mostly received with positive feedback. However, the initial promises of the game encouraged us to work further in future. 

\begin{figure}
	\centering
	\includegraphics[width=\columnwidth]{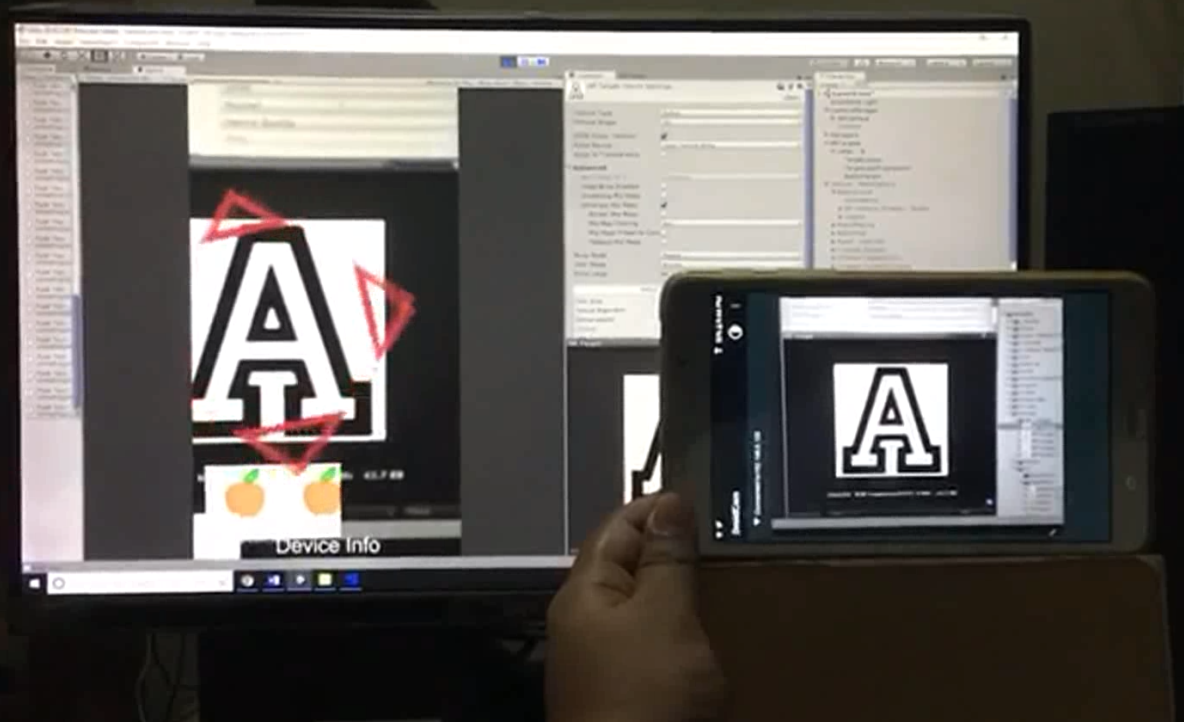}
	\caption{Applying VR to detect alphabets based on objects. \label{figVR}}
\end{figure}

\section{Conclusion}
In this paper, we presented {\methodName} a learning game application for special children. With our experiences with the children, we came to conclusion that they are more likely to accept the digital media as they find it more interesting than their real life toys and exercise elements. The benefits of engaging in digital media is helping them to their learning curve but removing intension of social communication as they feel more comfort of being alone. The teachers gave us an amazing point that if the digital media can also communicate, it will help them to encourage in social communication as well. We have already deployed $``$Vuforia$"$ (see Figure~\ref{figVR})  to inspire the user to detect letter from the real word, such as books or stickers to make them more interactive with environment. In future, we are planning to add virtual agents to help them with speech exercise.

	
\end{document}